\documentclass[twocolumn,showpacs,preprintnumbers,amsmath,amssymb,aps,prc]{revtex4-1}
	
\usepackage{xcolor}
\usepackage{verbatim}
\usepackage{graphicx}
\usepackage{dcolumn}
\usepackage{bm}
\usepackage{lineno}

\usepackage[colorlinks=true, allcolors=blue]{hyperref}
\begin{document}
\title{Backward-Angle ($u$-channel) Production at an Electron-Ion Collider}
\author{Daniel Cebra}
\affiliation{University of California, Davis}
\author{Xin Dong}
\affiliation{Lawrence Berkeley National Laboratory Berkeley CA}
\author{Yuanjing Ji}
\affiliation{Lawrence Berkeley National Laboratory Berkeley CA}
\author{Spencer R. Klein}
\affiliation{Lawrence Berkeley National Laboratory Berkeley CA}
\author{Zachary Sweger}
\affiliation{University of California, Davis}
\date{\today}

\begin{abstract}

In backward photoproduction of mesons, $\gamma p\rightarrow M p$, the target proton takes most of the photon momentum, while the produced meson recoils in the direction from which the photon came.  Thus the Mandelstam $u$ is small, while the squared momentum transfer $t$ is typically large, near the kinematic limit.  
In a collider geometry, backward production transfers the struck baryon by many units of rapidity, in a striking similarity to baryon stopping.  We explore this similarity, and point out the similarities between the Regge theories used to model baryon stopping with those that are used for backward production. 

We then explore how backward production can be explored at higher energies than are available at fixed target experiments, by studying production at an electron-ion collider.  We calculate the expected $ep$ cross sections and rates, finding that the rate for backward $\omega$ production is about 1/300 that of forward $\omega$s.  We discuss the kinematics of backward production and consider the detector requirements for experimental study. 

\end{abstract}

\maketitle

\section{Introduction}

In backward ($u-$channel) production, a photon interacts with a nucleon and produces a vector (or other) meson, albeit with kinematics that are reversed from the usual forward production.  In  conventional forward photoproduction, the meson takes most of the momentum of the photon, so the momentum transfer from the nuclear target ($\sqrt{t}$) is small.   In backward production, however, $t$ is large, but $u$ is small; the struck nucleon takes most of the momentum of the photon.  In the $\gamma p$ center-of-mass frame, the produced meson and the nucleon nearly switch places.  

In a collider geometry, in either ultra-peripheral collisions \cite{Klein:2020fmr,Contreras:2015dqa,Baltz:2007kq}
or in $ep/eA$ collisions, the target nucleon is shifted many units of rapidity, often ending up near mid-rapidity.  The produced meson takes most of the energy of the nucleon, ending up near the beam rapidity.  Seen in this light, backward production shares many similarities with baryon stopping in relativistic heavy-ion collisions.

Backward production has been studied using two different theoretical paradigms.  In the first, it is quantified using a set of Transition Distribution Amplitudes (TDAs) that are a kind of structure function for these reactions \cite{Pire:2015kxa}.   TDA models predict that the cross section for transversely polarized photons is much larger than that for longitudinally polarized photons.  Another feature of this paradigm is that for large $Q^2$ ($Q^2>|u|$), the cross section scales as $1/Q^8$.

\begin{figure*}
  \begin{center}
    \includegraphics[width=0.2\textwidth]{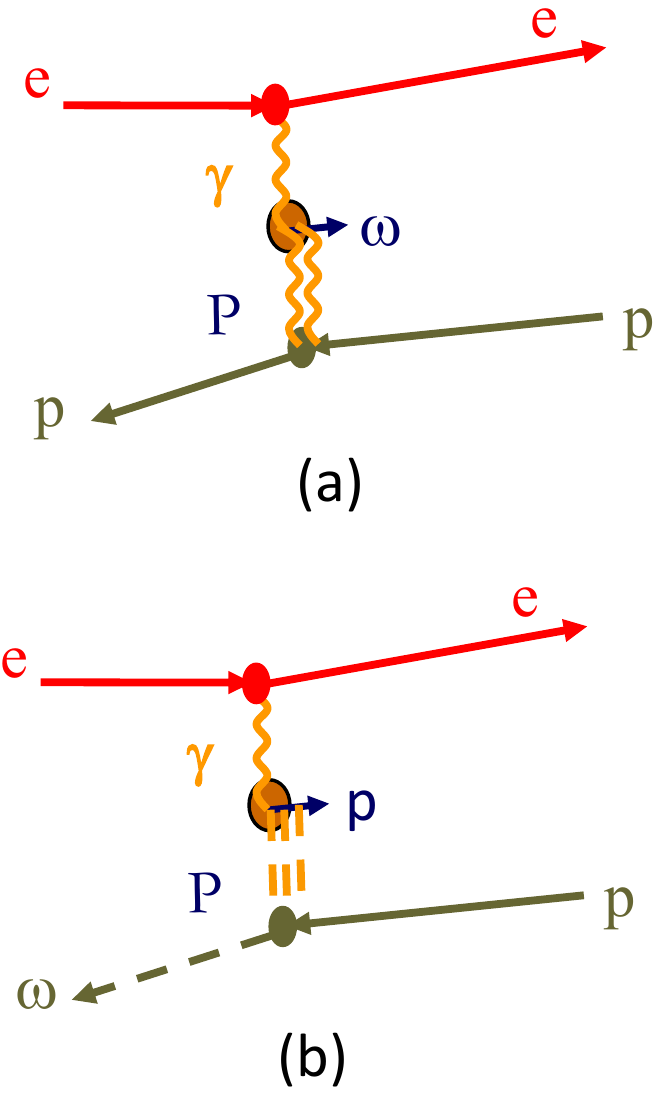}
        \includegraphics[width=0.5\textwidth]{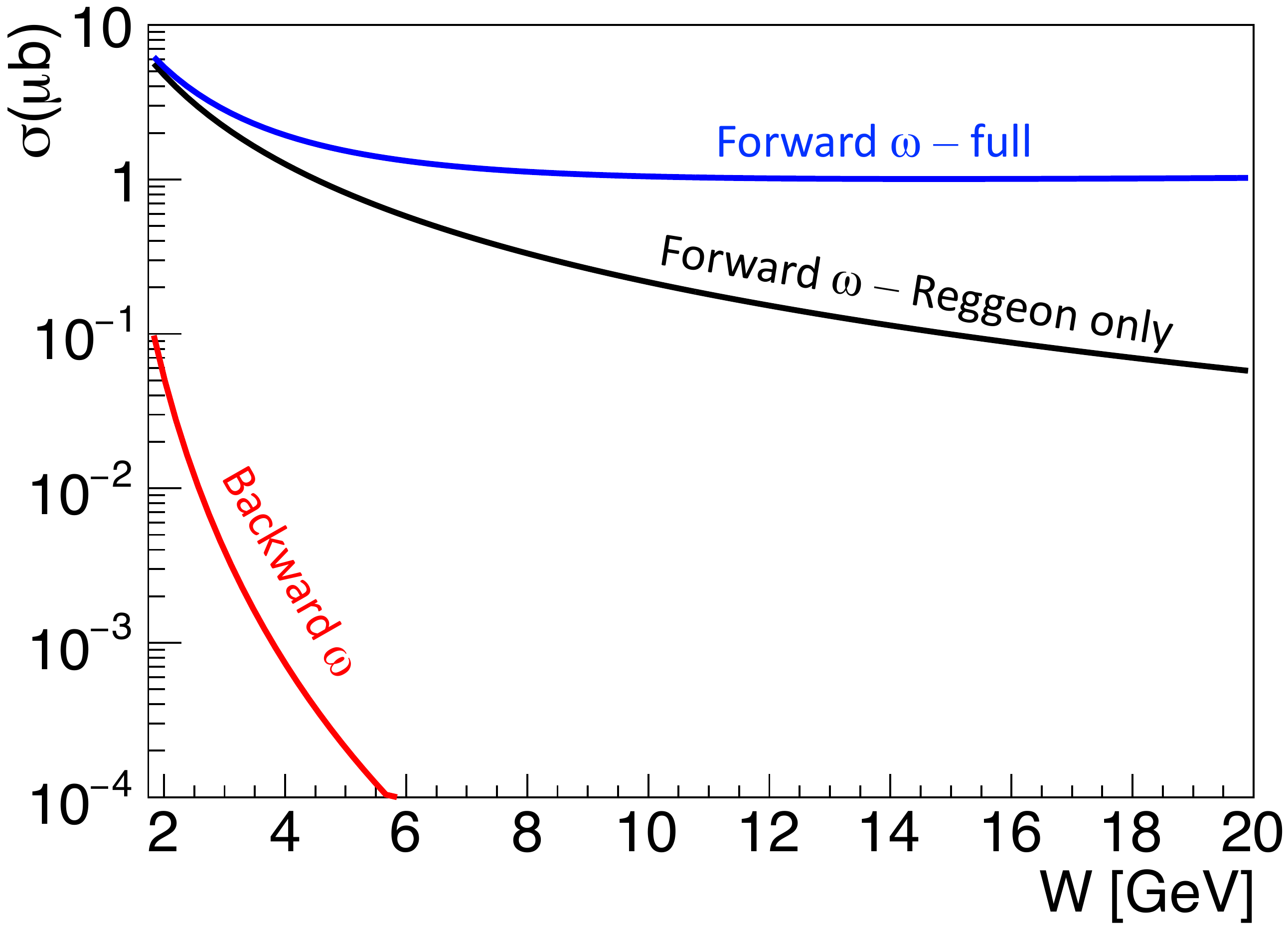}
    \caption{Schematic diagrams showing (a) forward $\omega$ production and (b) backward $\omega$ production.  In backward production, the $\omega$ and the proton essentially swap places (in rapidity, with respect to the $\gamma p$ center of mass).  The right-hand plot shows the cross sections as a function of $W$ for forward and backward $\omega$ production.}
    \label{fig:backwardproduction}
  \end{center}
\end{figure*}

The second paradigm is an extension of Regge phenomenology \cite{Clifft:1977yi,Laget:2021qwq}.  Figure \ref{fig:backwardproduction}(a) demonstrates conventional forward $\omega$ production via Reggeon exchange; the baryon momentum is little changed.  Figure \ref{fig:backwardproduction}(b) shows backward production via the exchange of a Reggeon, which also carries baryon number. The Reggeon carries the baryon number from the incident proton to the outgoing meson, obviating the need for it to transfer a large momentum.

Backward production has been studied at fixed-target experiments \cite{Gayoso:2021rzj,Guidal:1997by,Anderson:1969jw,Tompkins:1969jd}, but little is known about its behavior at higher energies.  The proposed U.S. electron-ion collider (EIC) will allow for the study of backward production of a variety of different mesons at different collision energies. It will also probe backward production with nuclear targets, and study high-energy electroproduction at large $Q^2$ \cite{Gayoso:2021rzj}. 

In Section II of this paper, we explore the connections between the Regge models used to model backward production and those used to study baryon stopping.  In Section III, we develop a detailed production model in $ep$ collisions.  We then examine the kinematics (Sections IV and V) and detector requirements (Section VI) to study backward $\omega$ and $\rho$ production.  The $\omega$ is among the most-studied for backward production. It decays to an all-neutral final state $\pi^0\gamma$.  We also consider an experimentally complementary channel, $\rho^0\rightarrow\pi^+\pi^-$, in order to study detection capabilities using charged-particle tracking.   We discuss the reaction kinematics, and the detector requirements to observe the final states from $\omega$ and $\rho^0$ decay, including the stopped proton.

We consider collisions at different EIC energies.  For some final states, the lower-energy-collision products may be more experimentally accessible than products from top EIC energies.  In Section VII, we also briefly comment on prospects at the proposed Chinese EiCC, which will collide 20 GeV protons with 3.5 GeV electrons \cite{Anderle:2021wcy}, and at the proposed high-energy LHeC \cite {LHeC:2020van}.

\section{Connections with Baryon Stopping}

Backward reactions shift nucleons by many units of rapidity.  This is conceptually similar to the large baryon stopping observed in heavy-ion collisions~\cite{Busza:1983rj}, which has been observed in ultra-relativistic $pp$ and $p$Au reactions. 

Different models have been proposed to explain baryon stopping.  In many models, the stopping involves the valence quarks, which interact, lose energy and are thus shifted in rapidity. In heavy-ion collisions, the valence quarks in an incident nucleon typically undergo many interactions with partons (or strings) in the target, losing energy in each interaction. This process may be described statistically, leading to rapidity loss predictions which match the RHIC data \cite{Hoelck:2020iei}.  Baryon stopping can also be described using a hadron transport approach \cite{Mohs:2019iee}.   In Monte Carlo implementations, the baryon is divided into a quark and a diquark, with the diquark perforce carrying the baryon number.
In these more traditional stopping models, baryon stopping is connected with the loss of baryon momentum, since the valence quarks together carry a good fraction of the baryon momentum.

Other models describe baryon stopping using a baryon junction model.  In these models,  \cite{ToporPop:2004lve}, the three valence quarks in a baryon are connected together with three gluon flux lines that meet at a central junction. Baryon stopping is decoupled from the valence quarks, and thus is less tied to baryon energy loss. As Kharzeev noted, ``contrary to a widely accepted belief, the traces of baryon number in a high-energy process can reside in a non-perturbative configuration of gluon fields, rather than in the valence quarks'' \cite{Kharzeev:1996sq}.  This is a soft phenomenon, and therefore is not visible in the hard collisions that are usually used to probe baryons.  However, the baryon junction can be easily shifted several units of rapidity, producing significant baryon-number stopping. The model has been used to explain the observed stopping in PbPb collisions at the CERN SPS \cite{Vance:1998vh}.  This model predicts that baryon stopping should go as $\sigma\propto \cosh(y/2)\approx \exp(-y/2)$.

The baryon-junction model has theoretical underpinnings in Regge theory.  The baryon-junction wave function is a ``quarkless closed string configuration composed from a junction and an antijunction.  In the topological expansion scheme, the states lie on a Regge trajectory'' \cite{Kharzeev:1996sq}, with the baryon intercept at zero and the Reggeon intercept at 1/2. 

In hadron-hadron collisions, including heavy-ion collisions, the Reggeon can interact with an incident baryon via a double-Reggeon interaction, or via a Reggeon-Pomeron interaction.  In the former case, both baryons are stopped, while in the latter only one is stopped.  The fraction of interactions with large stopping can be estimated; for lead-lead collisions at the LHC, the fraction is about 5\% \cite{Kharzeev:1996sq}.  A later calculation found a stopping cross section of 1-2 mb \cite{ToporPop:2004lve}, comparable to the 5\% fraction.

The 5\% fraction for stopping interactions in heavy-ion collisions is one to two orders of magnitude higher than the fraction of backward-production reactions when compared to the total (backward + forward) cross section in $ep$ collisions.  This is unsurprising; in heavy-ion collisions, there are very high densities of gluons at low Bjorken$-x$ with which the Reggeon can interact, while the $q\overline q$ dipole is accompanied by a much smaller gluon cloud.  

This picture predicts that there should be significant baryon stopping in $pp$ collisions, via reactions like $pp\rightarrow pp\omega$, mediated by a Reggeon-Pomeron exchange.  One of the protons loses only a small amount of momentum while the other proton is significantly stopped.  The meson takes most of the momentum lost by the stopped proton.   Per \cite{Kharzeev:1996sq}, there are occasional double-stopping reactions, with both protons at mid-rapidity, with mesons produced at large rapidity carrying away most of the excess momentum and energy.  However, ALICE data on the $\overline p:p$ ratio at midrapidity in $pp$ collisions is consistent with the the conventional stopping predictions, and below that predicted by HIJING/B, which implements a baryon-junction model. 

The baryon-junction model could also be used to predict baryon stopping in general photonuclear reactions, $\gamma p/A\rightarrow X$ collisions, which are being studied by STAR at RHIC by measuring the net-baryon rapidity \cite{Prithwish2022}.   

From these reactions, it is a small step further to exclusive $\gamma p$ reactions, such as vector-meson production. Here, a clear bifurcation is expected, with minimal baryon stopping in forward production, but with large stopping in backward production.  It will be interesting to see if more complex nonexclusive photonuclear interactions exhibit a similar bifurcation.

Before continuing, it is worth mentioning an additional aspect of baryon stopping: charge stopping.   In exclusive backward photoproduction, either the incident baryon and the final state meson share a valence quark, or they don't.  If they do share a quark, then only two quarks need to be decelerated, but if they do not share a quark, then three quarks must be stopped.   Decelerated charges may emit bremsstrahlung radiation; the total infrared radiation is sensitive to the number of charges and their acceleration \cite{Low:1958sn}.  Thus, a measurement of soft photon radiation may probe the manner in which the initial proton and final state meson share quark content.    One test would be to compare this soft radiation for $\rho$ or $\omega$ backward production with that for the $\phi$, where the meson shares no valence quarks with the incident baryon.   
These exclusive backward production reactions should provide much information on baryon stopping.  In the following sections, we discuss how they may be studied at the EIC. 

\section{Backward photoproduction and electroproduction model}
\label{section:crosssections}

We use a Regge-based approach to model backward production, since it requires a relatively small number of free parameters.   These parameters are determined from fixed-target backward photoproduction experiments over a limited range of photon energies.   The extrapolation required to reach EIC energies introduces uncertainty in the total rate and rapidity distributions, but does not affect the qualitative conclusions in the paper. 

In Regge models, the cross section for photoproduction as a function of photon energy $k$ is 
\cite{Crittenden:1997yz,Klein:1999qj}
\begin{equation}
    \sigma_{\gamma p\rightarrow Vp}(k) = XW^\epsilon + YW^{-\eta},
    \label{eq:sigma}
\end{equation} 
where W is the $\gamma p$ center-of-mass energy in GeV, and $X$, $Y$, $\epsilon$ and $\eta$ are parameters determined from fits to data.  The first ($X$,$\epsilon$) term represents Pomeron exchange. We will focus on the second term, modeling Reggeon exchange, since this is more directly comparable to backward production.  Reggeon exchange admits a much wider range of quantum numbers, including charge \cite{Klein:2019avl}. For the $\omega$, a fit to existing data found $\epsilon=0.22$, $X=0.5 \;\mu$b,
$\eta=1.92$ and $Y=18 \; \mu$b \cite{Crittenden:1997yz,Klein:1999qj}. 

This parameterization does not account for near-threshold behavior.  The threshold region is very narrow, with the forward cross section maximum occurring around $W=W_{\rm max}=1.8$ GeV \cite{Strakovsky:2014wja}, close to the kinematic threshold of $W_T=m_p+m_\omega=$ 1.74 GeV.   We account for this region by adding a linear scaling factor $T(W)$ to  Eq. \ref{eq:sigma} for $W< W_{\rm max}$:
\begin{equation}
    T(W)=   {\rm Min}\Big(\frac{W-W_T}{W_{\rm max}-W_T},1.0\Big).
   \label{eq:sigmathreshold}
\end{equation}
The $t$ dependence for forward production is usually given by an exponential,
\begin{equation}
    \frac{d\sigma}{dt}\sim\exp{(-Bt)}.
    \label{eq:dsdt}
\end{equation}
For the $\omega$, $B=10 \pm 1.2 ({\rm stat.}) \pm 1.3 ({\rm syst.})$ GeV$^{-2}$ \cite{ZEUS:1996zse}.

For backward production, we combine Eqs. \ref{eq:sigma}, \ref{eq:sigmathreshold} and \ref{eq:dsdt} and swap Mandelstam $u$ for $t$:
\begin{equation}
    \sigma_{\gamma p\rightarrow Vp}(k) = AW^{-\eta} T(W) \exp{(-Cu)}.
    \label{eq:sigmafull}
\end{equation}
The parameters are determined from a fit to backward production data.

There are some challenges to fitting backward photoproduction data.  At threshold, both $t$ and $u$ are small, and forward and backward production are essentially indistinguishable.  At slightly higher energies, baryon resonances that decay to $\omega p$ can contribute to the cross section, as was seen in a study of backward photoproduction using 1.5 to 3.0 GeV photons at the LEPS synchrotron  \cite{Morino:2013raa}. 

To minimize these obstacles, we focus on higher-energy backward $\omega$ photoproduction data from Daresbury Laboratory's NINA 5 GeV electron synchrotron.  The Daresbury group measured cross sections for photon energies from 2.8 to 4.8 GeV \cite{Clifft:1977yi}. These data are also discussed in detail in Ref. \cite{Sibirtsev:2002at}.  A fit to these measurements found, as a function of photon energy ($k$) in the target frame:
\begin{equation}
    \frac{d\sigma}{du}\Big|_{u=0} = A \Big(\frac{k}{\rm 1 \;GeV}\Big)^{-\eta}
     = A \Bigg(\frac{W^2-m_p^2}{2m_p(\rm 1 \;GeV)}\Bigg)^{-\eta},
\end{equation}
where $A=4.5\pm 1.0 \; \mu$b/GeV$^2$ and $\eta=2.7\pm0.2$.

The $u-$dependence of the cross section is also found in Ref. \cite{Clifft:1977yi}.  Parameterization of $\sigma(\gamma p\rightarrow \omega p)$ is expressed as the sum of two amplitudes,
\begin{equation}
    \frac{d\sigma(s,u)}{du}= \bigg|A(u)s^{\alpha(u)-1} + B(u)e^{i\phi(u)}s^{-n/2}\bigg|^2.
\end{equation}
The first term represents Reggeon exchange and the second term denotes parton exchange, for which $n=8$ is taken. They assume that the two components do not interfere ({\it i. e.} $\phi=90^0$), so the two components are squared separately and added.  The parton-exchange component drops very rapidly with increasing $s$, so we focus on the Reggeon-exchange component.   

The Daresbury group's Fig. 4 shows that $A(u)$ drops off rapidly with increasing $u$, reaches a minimum for $u\approx 0.15$ GeV$^2$, rises, and then drops off again.  This behavior is quite different from forward production, where a single exponential, Eq. \ref{eq:dsdt}, describes the data well. 

In this paper, we will focus on the rapid drop off, which contains much of the cross section.  We fit the 4.7 GeV data in the Daresbury group's Fig. 2b to $d\sigma/du = (100. \pm 3.) \exp{(-20.8 \pm 1.2\ {\rm GeV}^{-2} u)}$ mb/GeV$^2$.  A comparison of their 3.5 and 4.7 GeV data shows that the lower-energy set has a larger component at large $u$.  As their Fig. 4 and text show, this is associated with the parton-exchange component, so we will not consider the high $|u|$ data here.  Uncertainties on the energy-dependence of the cross section will largely impact the spread in rapidity of the final-state proton, while variation in pseudorapidity ($\eta$, henceforth) will largely affect the $p_T$ distribution of the vector meson as well as the rapidity of the final-state proton.

The polarization of the produced mesons introduces another uncertainty.  Forward production of vector mesons follows $s$-channel helicity conservation (SCHC), where produced mesons have the same spin as the incident photons.   The Reggeon in Fig. \ref{fig:backwardproduction}(b) that transfers baryon number likely also carries spin.   Here, we will assume that the produced mesons are unpolarized, but will later quantify the uncertainty that this introduces, which is not large. 

Figure \ref{fig:backwardproduction} (right) compares the forward and backward photoproduction cross sections.  At the production maximum, backward production is about 1\% of the forward-production cross section, but the fraction decreases with increasing $W$.

The EIC will also probe backward electroproduction, with photon $Q^2>0$.  An experiment at Jefferson Lab studied backward $\omega$ electroproduction at two $Q^2$: 1.6 GeV$^2$ and 2.45 GeV$^2$, at fixed $W=2.21$ GeV \cite{JeffersonLabFp:2019gpp}. The cross section for longitudinally polarized photons, $\sigma_L$ dropped rapidly with increasing $Q^2$, roughly as $1/Q^{10.22}$, while the cross section for transversely polarized photons, $\sigma_T$ was much less sensitive to $Q^2$, falling as $1/Q^{1.08}$.  The strong $Q^2$ dependence of $\sigma_L $\ cannot continue as $Q^2\rightarrow 0$, since it would diverge; this also holds for the $1/Q^8$ dependence predicted in the TDA approach.  We model electroproduction by assuming that backward $\omega$ production has the same $Q^2$ dependence as forward production \cite{Klein:2019avl}:
\begin{equation}
\sigma(Q^2,W) = \sigma(0,W)\bigg( \frac{M_\omega^2}{M_\omega^2+Q^2}\bigg)^{2.09+0.73(M_\omega^2+Q^2)/{\rm GeV}^2}.
\end{equation}
The coefficients in the power law were obtained from a fit to $\rho^0$ production data since no similar fit is available for the $\omega$ \cite{H1:2009cml}.  The corresponding coefficients for the $\phi$ are very similar, so this assumption should be quite accurate, at least for forward production.

Backward $\rho^0$ production has been less studied than the $\omega$.  However, comparisons of $\rho^0$ \cite{CLAS:2001zxv} and $\omega$ photoproduction \cite{Clifft:1977yi} at large momentum transfer (large $|t|$) indicate that their backward production cross sections are similar \footnote{See Fig. 13 of Ref. \cite{Gayoso:2021rzj}.}, even though the $\rho^0$ forward production cross section is ten times greater. 

The $u$-dependence favors smaller values of $u$ than is seen for $t$ in forward production, so the typical $p_T$ of the $\omega$ and proton are smaller than in forward production. To the extent that this represents a source size, the apparent source is larger for backward production than for forward production.

\section{Simulating Backward Production}
\begin{figure*}
  \begin{center}
    \includegraphics[width=0.9\textwidth]{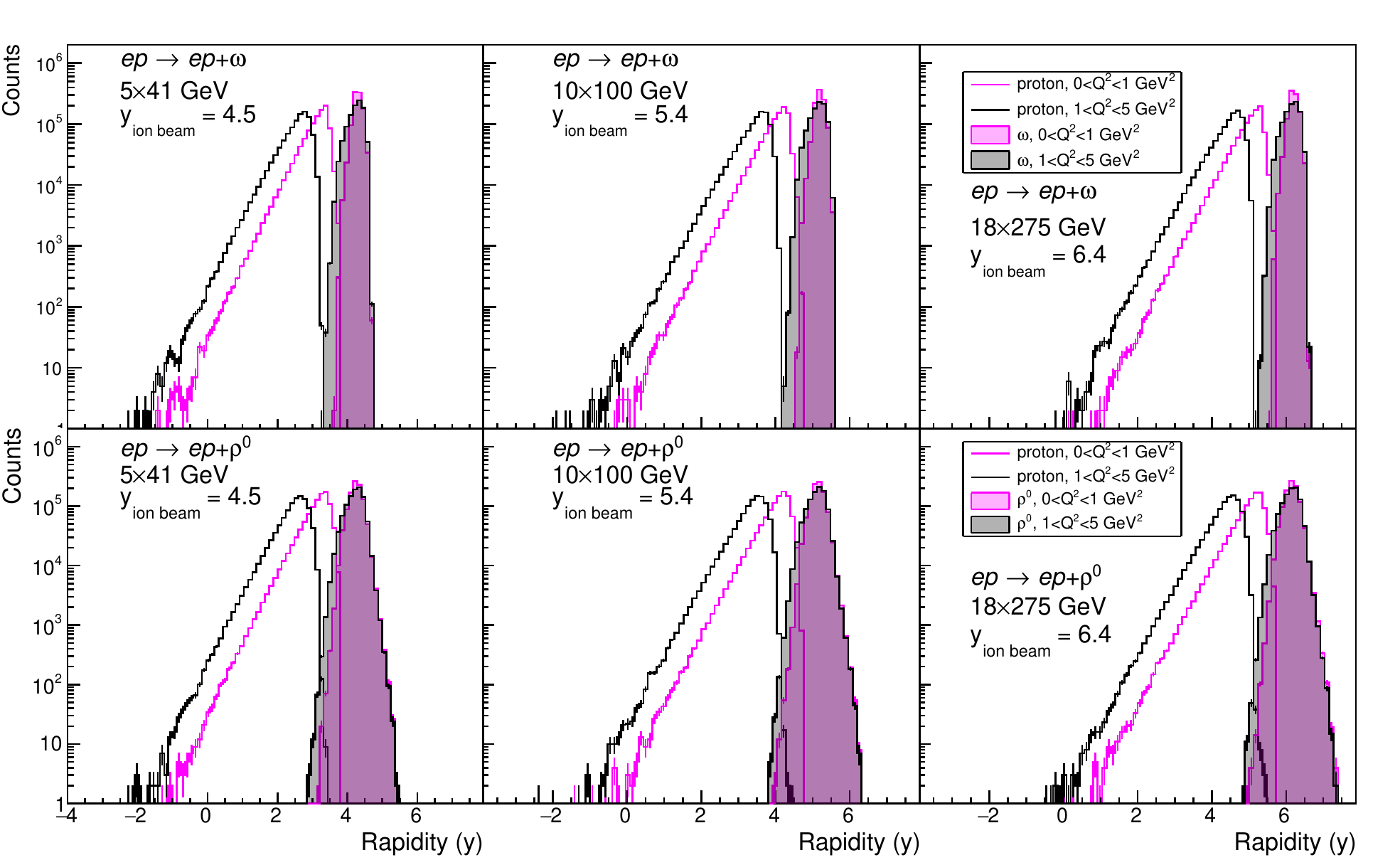}
    \caption{The proton and meson rapidities for $\omega$ (top row) and $\rho^0$ (bottom row) $u$-channel photoproduction and electroproduction at several EIC energies.  The magenta distributions are for photoproduction, while the black are for electroproduction.  The shaded histograms show the vector meson, while the unshaded ones are for the scattered proton.  The non-zero photon $Q^2$ shifts both distributions slightly toward negative rapidity.}
    \label{fig:rapiditycomparison}
  \end{center}
\end{figure*}
\subsection{Production Kinematics}

Calculations were done using the eSTARlight Monte Carlo generator \cite{Lomnitz:2018juf}, which was written to simulate forward vector-meson production. It models $ep$ and $eA$ collisions with intermediary photons with different virtuality. The code is available on GitHub \cite{eSTARlight}.  eSTARlight has an interface similar to STARlight \cite{Klein:2016yzr}, which is used to simulate ultra-peripheral collisions, but has a very different internal structure.  For this work, we added code to eSTARlight to model backward production.

The $u$-channel calculations follow the standard $t$-channel production calculations. First, the photon energy and $Q^2$ are sampled according to the cross sections described in Sec. III of Ref. \cite{Lomnitz:2018juf}. Then the mass of the produced meson is sampled from a Breit-Wigner distribution. 

In the photon-proton center-of-mass frame, the  vector meson energy is
\begin{equation}
    E_{V} = E_\gamma+\frac{m_V^2 + Q^2}{2(E_\gamma+E_p)}.
    \label{equation:energycondition}
\end{equation}
With this constraint, the energy and momenta are calculated for the vector meson and scattered proton in the $t$-channel. 
The $u$-channel kinematics are obtained by swapping the scattered proton and vector meson momenta in the proton-meson (proton-photon) center-of-mass frame. In this frame, the masses and energies of the final-state particles are unchanged by the reversal, but the boost back to the lab frame produces a vector meson with a large rapidity.

We will consider three different EIC energies: 275 GeV protons on 18 GeV electrons, 100 GeV protons on 10 GeV electrons, and 41 GeV protons on 5 GeV electrons.  The choice of beam energies will have significant implications for detection efficiency as will be demonstrated in Sec. \ref{section:detection}.

\subsection{Final-State Particle Kinematics}

In deep inelastic scattering the angle of the scattered electron is
\begin{equation}
\text{cos}(\theta_e) = 1 - \frac{Q^2}{2E_eE_{e'}},
\end{equation}
where $e$ is the incident beam electron and $e'$ is the scattered electron.

The scattered proton kinematics are constrained by the energy condition in Eq. \ref{equation:energycondition}. The vector-meson daughter distribution is calculated by boosting to the rest-frame of the vector meson and calculating the magnitude of the daughter momenta. The direction of the daughter momenta depends on the polarization of the vector meson. Per SCHC, vector mesons produced via the $t$-channel mechanism retain the spin polarization of their virtual photon. 

By contrast, $u$-channel production involves a baryon-exchange trajectory, so the vector meson is unlikely to retain the photon's polarization. Therefore, simulations of unpolarized vector mesons are used in this paper.  However, as Sec. \ref{subsection:rhoproduction} will show, the polarization does not have a large effect on the detection efficiency. Though its effects on efficiency are small, measuring final-state polarization will be important in understanding backward production at the EIC.

\section{Rates and Kinematics for Backward Production}

We consider two complementary channels: the all-neutral decay $\omega\rightarrow\pi^0\gamma\rightarrow\gamma\gamma\gamma$, which can be studied using only calorimetry; and the all-charged decay $\rho\rightarrow\pi^+\pi^-$, where the $\pi^\pm$ can be reconstructed in tracking detectors.   The kinematics of backward $\rho$ production are unknown, but they are likely to be similar to that for the $\omega$. There is a 1.5\% branching ratio for $\omega\rightarrow\pi^+\pi^-$, but this channel will mostly be visible via its interference with $\rho\rightarrow\pi^+\pi^-$ \cite{STAR:2017enh}.

Table \ref{tab:rates} shows the $\omega$ cross sections and production rates (not including branching ratios) for five different collision energies for an integrated luminosity of 10 fb$^{-1}$.  The backward-production cross sections are about 1/300 that of the forward production channel.   Although this is a small fraction, the rates are large enough that millions of events per year are expected at all five collision energies.  For the high-luminosity configuration with 10 GeV electrons, the rates could reach 100 million events/year. 

The rapidities of the proton and meson from electroproduction and photoproduction of $\omega$ and $\rho^0$ mesons are shown in Fig. \ref{fig:rapiditycomparison}.  The vector mesons are at fairly large rapidity, while the proton is distributed over a wide rapidity range.  The configuration with the proton at large rapidity, near the vector meson, corresponds to the smallest $W$.  The diminishing $d\sigma/dy_{\rm proton}$ with decreasing $y$ reflects the decrease in the cross section with increasing $W$.  The simulated $dN/dy$ is consistent with an exponential dependence on rapidity, but it is considerably steeper than the previously discussed $\exp(-y/2)$ stopping in heavy-ion collisions \cite{Vance:1998vh}.   If the cross-section dropped more slowly with increasing $W$, then the agreement would be better.  This slope is a fairly direct measure of $\eta$ in Eq.~(4). 

As the proton beam energy increases, the vector-meson rapidity distribution shifts to larger rapidity.  The detection efficiency is very sensitive to the vector-meson rapidity, so the beam energy has a significant effect on the detector acceptance. In contrast, the proton rapidity distribution is less sensitive to the beam energy.

\begin{figure}
  \begin{center}
    \includegraphics[width=0.49\textwidth]{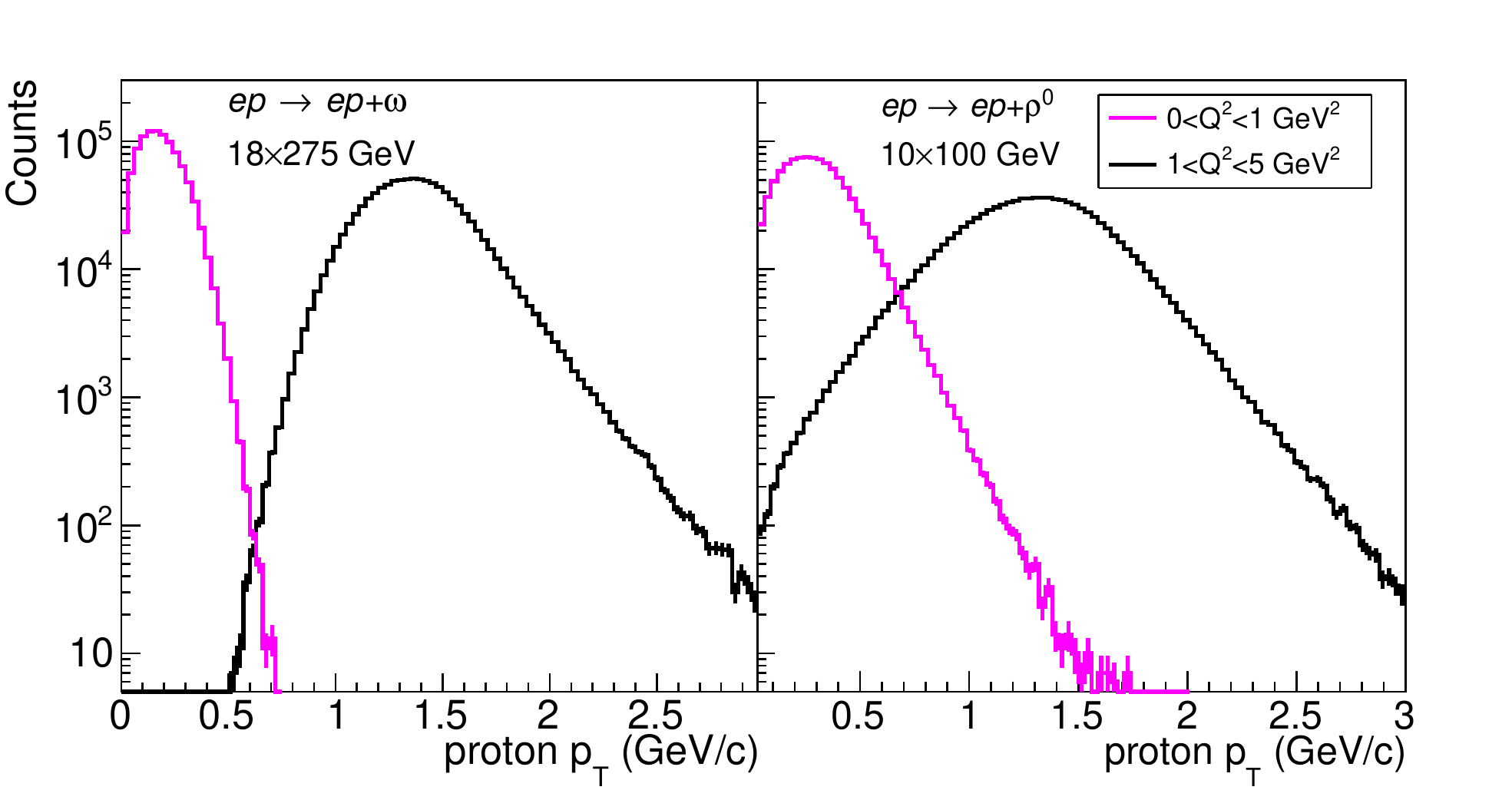}
    \caption{Proton $p_T$ spectra for $u$-channel $\omega$ and $\rho^0$ electroproduction (black) and photoproduction (magenta).}
    \label{fig:protonpT}
  \end{center}
\end{figure}

The $p_T$ spectra of scattered protons in electro/photoproduction are plotted in Fig. \ref{fig:protonpT}.  The low average $p_T$ of protons in photoproduction may make proton identification more challenging than in electroproduction.

\begin{figure}
  \begin{center}
    \includegraphics[width=0.5\textwidth]{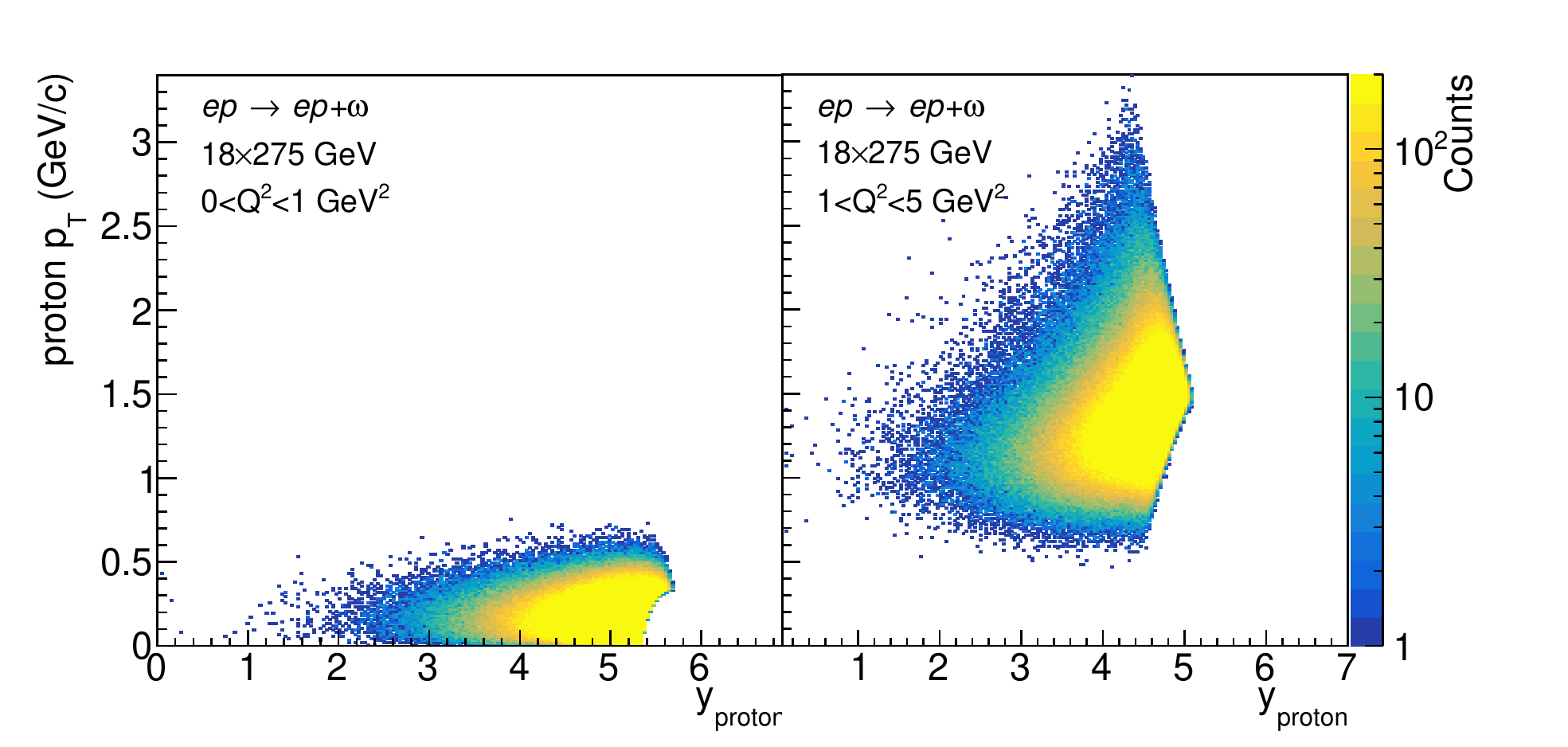}
    \caption{Proton $p_T$ for $u$-channel  $\omega$ electroproduction (black) and photoproduction (magenta) as a function of the proton rapidity.}
    \label{fig:protonpTVsY}
  \end{center}
\end{figure}

Figure \ref{fig:protonpTVsY} shows how the proton $p_T$ spectrum depends on rapidity, for both photoproduction and electroproduction.  For photoproduction, the $p_T$ depends only slightly on the rapidity.  For electroproduction the mean proton $p_T$ is largest at forward rapidities, diminishing slowly as rapidity decreases. 

\begin{figure}[tb]
  \begin{center}
    \includegraphics[width=0.45\textwidth]{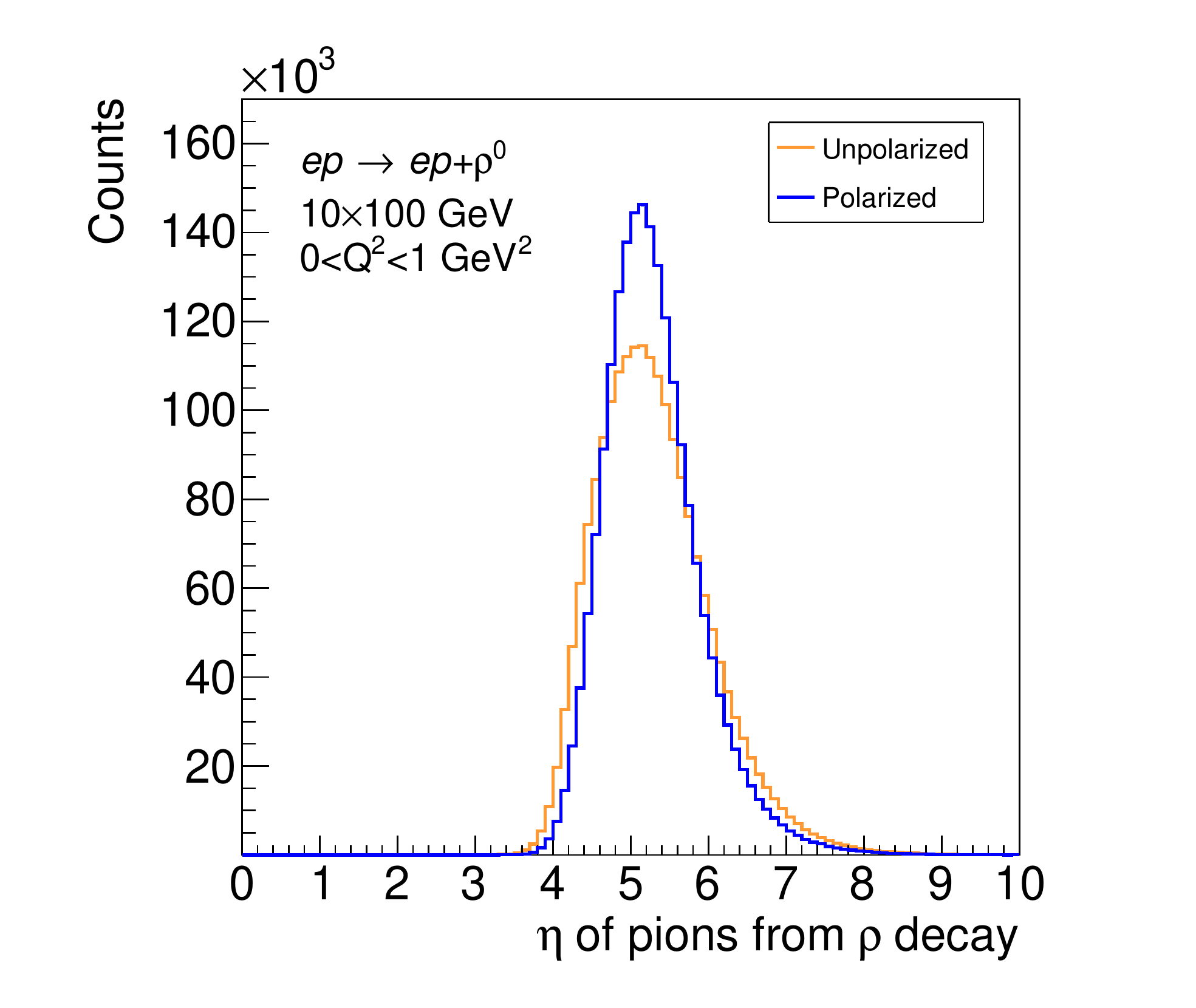}
    \caption{Pion pseudorapidity distributions from polarized (red) and unpolarized (blue) $u$-channel $\rho$ photoproduction at 10$\times$100 GeV.}
    \label{fig:polarization}
  \end{center}
\end{figure}

The pseudorapidities of the daughter particles from the vector meson decays depends on whether the decays are polarized or unpolarized.  Figure \ref{fig:polarization} compares pseudorapidity distribution for $\pi^\pm$ from $\rho^0$ decay for the two cases.  The distributions have almost the same mean, but the distribution is slightly wider in the unpolarized case.  This will not have a large effect on the detection efficiency.  In the remainder of this work, we will assume that the decays are unpolarized.

\begin{figure}
  \begin{center}
    \includegraphics[width=0.49\textwidth]{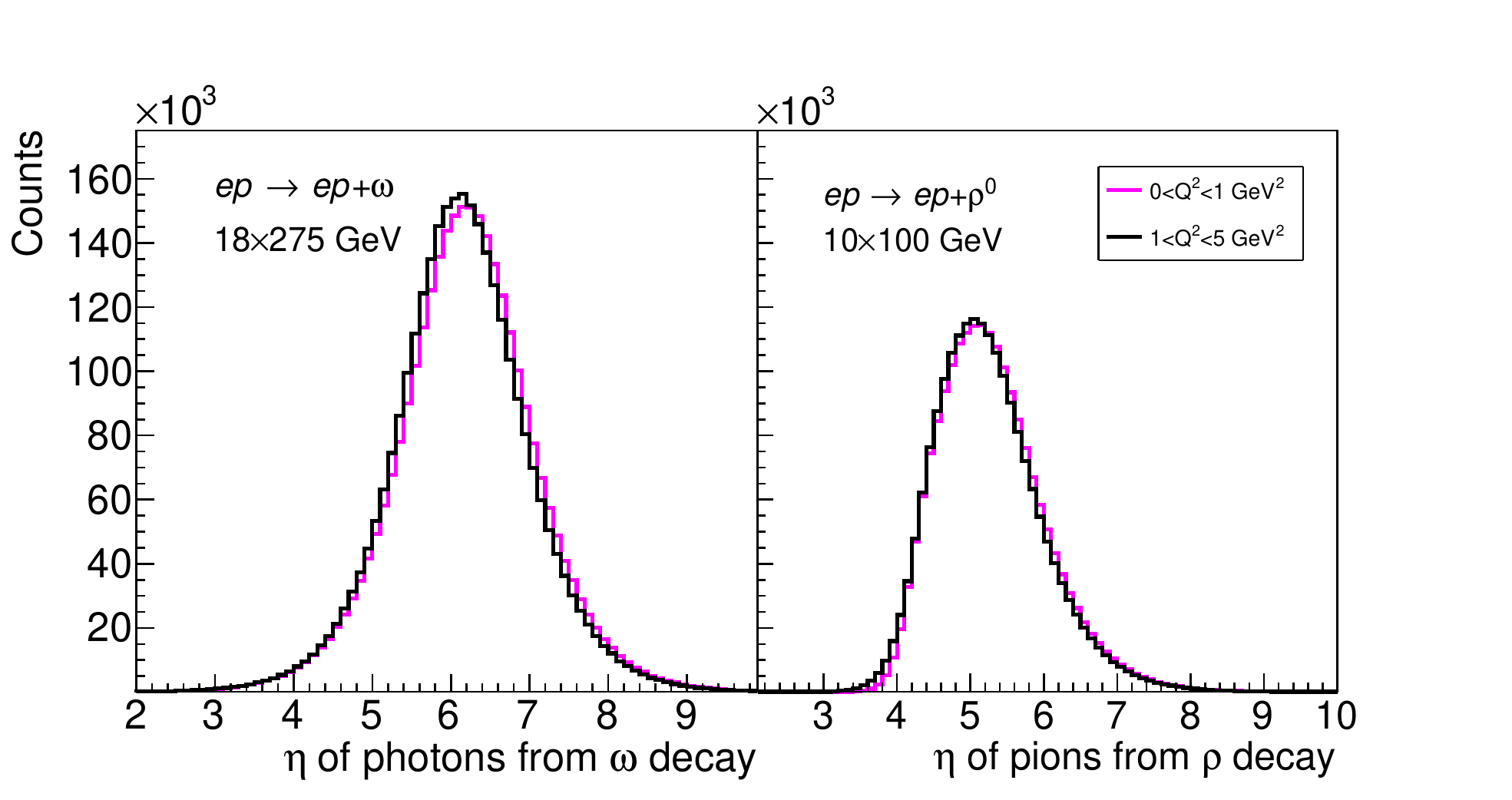}
    \caption{Vector-meson decay daughter pseudorapidity  distributions for $u$-channel $\omega\rightarrow\pi^0\gamma$ and $\rho^0\rightarrow\pi^+\pi^-$ electroproduction (black) and photoproduction (magenta).}
    \label{fig:photovselectro}
  \end{center}
\end{figure}

Figure \ref{fig:photovselectro} shows the final-state particle photons from $\omega$ decay and $\pi^\pm$ from $\rho$ decay, for photoproduction and electroproduction.  The photon $Q^2$ has little effect on the distributions. 

\begin{table*}
\begin{tabular}{| c  c c  c c |}
\hline
{Collider}&{$t$-channel}&{$u$-channel}&{     }&{$u$-channel events}\\
{energy}&{$\sigma_\text{Tot}$ (nb)}&{$\sigma_\text{Tot}$ (nb)}&{     }&{per 10 fb$^{-1}$}\\
\hline 
5$\times$41 GeV & 501 & 1.8 & & 1.8$\times$10$^7$  \\
5$\times$100 GeV & 583 & 1.9 & & 1.9$\times$10$^7$  \\
10$\times$100 GeV & 651 & 2.0 & & 2.0$\times$10$^7$ \\
10$\times$275 GeV & 758 & 2.2 & & 2.2$\times$10$^7$  \\
18$\times$275 GeV & 825 & 3.2 & & 3.2$\times$10$^7$  \\
\hline
\end{tabular}
\caption{The $\omega$ cross sections for forward and backward production and total number of backward $\omega$ events per 10 fb$^{-1}$ of integrated luminosity. The $t-$channel cross section includes both Pomeron and Reggeon exchange.  The rates are summed over all $\omega$ final states.}
\label{tab:rates}
\end{table*}

\section{Detection of Backward-Production Events}
\label{section:detection}
\subsection{Model Detector}

We will study detection efficiencies for a detector like the reference detector described in the EIC Yellow Report~\cite{AbdulKhalek:2021gbh}.  The reference detector was the starting-point for the ATHENA, ECCE, and CORE proposals.  The acceptance is divided into three regions shown in Tab. \ref{tab:detector}: a central detector, the B0 detector and the far-forward detectors, of which the Zero-Degree Calorimeter is relevant here.

\begin{table*}
\begin{tabular}{| l l c |}
\hline
{{\textbf{Detector}}}&{\textbf{Capabilities}}&{\textbf{Pseudorapidity Coverage}}\\
\hline
{Central}&{charged-particle tracking}&{$-3.5<\eta<3.5$}\\
{ }&{electromagnetic calorimetry}&{ }\\
{B0}&{charged-particle tracking}&{$4.6<\eta<5.9$}\\
{ }&{potential electromagnetic calorimetry}&{ }\\
{ZDC}&{electromagnetic calorimetry}&{$\eta>\sim6.1$}\\
\hline
\end{tabular}
\caption{The different regions and assumed pseudorapidity coverage per the 
EIC Yellow Report~\cite{AbdulKhalek:2021gbh}.}
\label{tab:detector}
\end{table*}

The central detector covers the region $-3.5<\eta<3.5$, with charged-particle tracking and electromagnetic and hadronic calorimeters.  For backward production, the central detector is primarily used to detect the scattered proton.

The B0 detector fits inside one of the accelerator magnets.  In the reference design, it included a charged particle tracking system, covering approximately $4.6<\eta<5.9$.  If physical space permits, it might also include an electromagnetic calorimeter.  In this paper, we consider the possibility of electromagnetic calorimetry in the B0 system in order to demonstrate how much additional physics it enables

The far-forward detector includes a zero-degree calorimeter (ZDC) which covers pseudorapidities above about 5.9 to 6.2.  Here, it is taken to cover $\eta>6.1$.

It may also be useful to tag the scattered electrons, in either the central detector (at large $Q^2$) or in the low-$Q^2$ tagger.   At very low $Q^2$, the electron may remain in the beampipe, and so not be visible; this did not prevent photoproduction studies at HERA.  Since the scattered electron has no significant difference in forward or backward scattering, we do not discuss this further. 

One complication in EIC detectors is that the beams do not meet head-on, but have a 25 mrad (or 35 mrad depending on IR) crossing angle.  In the frame of reference where the detector is aligned along the proton/ion beam axis, this has the effect of giving the final state a small transverse Lorentz boost, while in the electron-beam frame, the acceptance in pseudorapidity varies slightly with the azimuthal angle.  We ignore this small effect here. 

\subsection{$\omega\rightarrow\pi^0\gamma$}

\begin{figure*}
  \begin{center}
    \includegraphics[width=1.0\textwidth]{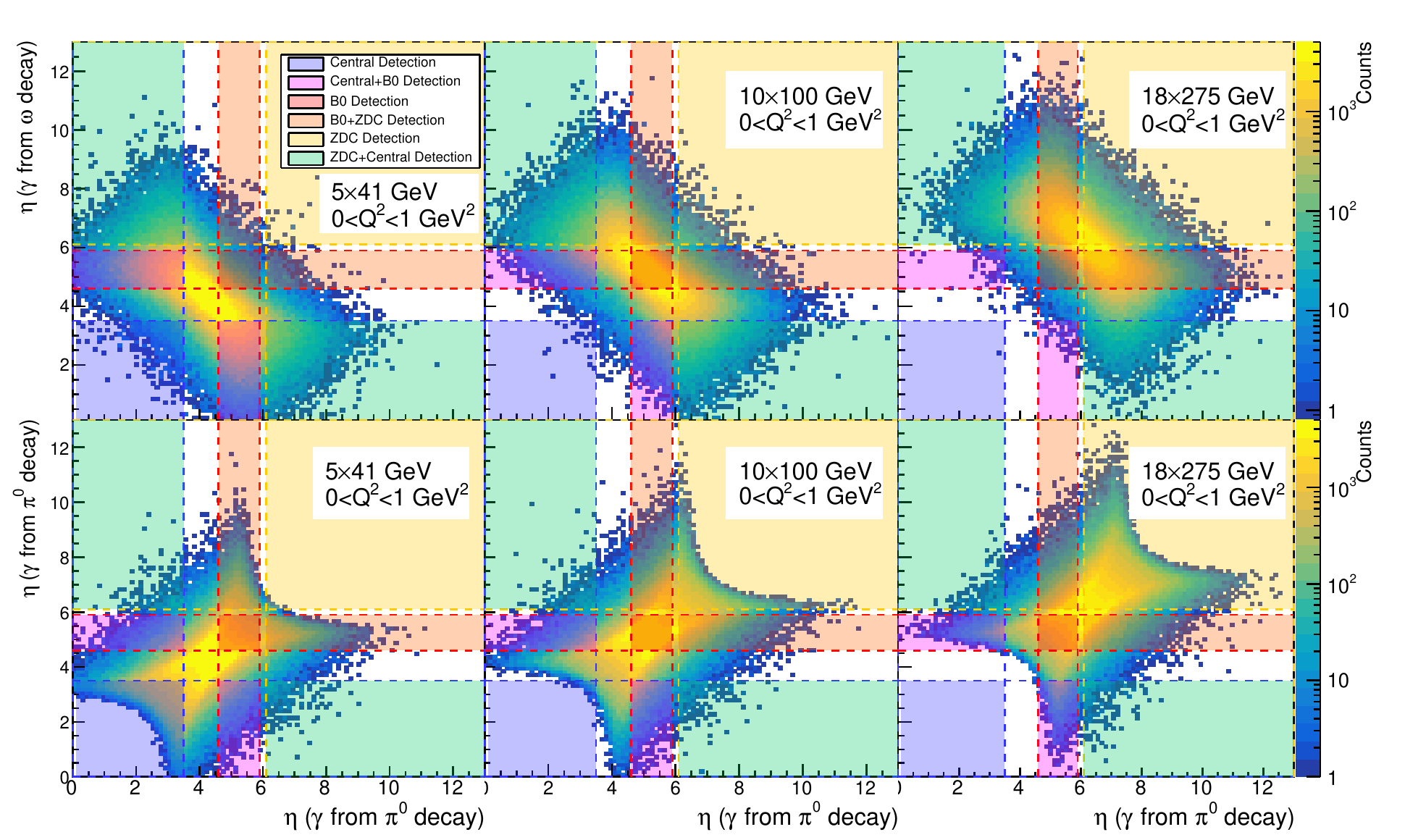}
\caption{Pseudorapidity distributions of photons from $u$-channel $\omega$ production at three EIC $ep$ collision energies. The shaded colored regions show the acceptance in pseudorapidity of the central, B0 and forward detectors.  The top row 2-d scatter plots the pseudorapidity of the photon from the $\omega$ decay and one of the photons from the $\pi^0$, while the bottom row shows the pseudorapidities of the two photons from the daughter $\pi^0$.}
    \label{fig:omegaenergyscan}
  \end{center}
\end{figure*}

\begin{table}[b]
\begin{tabular}{| c c c c c c c |}
\hline
{Proton}&{  }&{$\omega$ eff.}&{  }&{$\omega$ eff.}&{  }&{$\rho$ eff.}\\
{beam energy}&{  }&{cent.+ZDC}&{  }&{cent.+B0+ZDC}&{  }&{cent.+B0}\\
\hline 
41 GeV &  & 1.4\% &  & 18\% &  & 13\% \\
100 GeV &  & 1.3\% &  & 41\% &  & 49\% \\
275 GeV &  & 6\% &  & 63\% &  & 0.7\% \\
\hline
\end{tabular}
\caption{Geometric efficiencies for vector-meson reconstruction for the detector reference design.}
\label{tab:geoeff}
\end{table}
 
The most common decay for the $\omega$ is $\omega\rightarrow\pi^+\pi^-\pi^0\rightarrow\pi^+\pi^-\gamma\gamma$, with a branching ratio of 89.2\%.  Because the final state contains four particles, it is not a preferred method for studying $u$-channel $\omega$ production.  Instead, we consider the next most common decay, $\omega\rightarrow\pi^0\gamma\rightarrow\gamma\gamma\gamma$, with a branching ratio of 8.3\%. It can be fully reconstructed in the ZDC, and, if built, the B0 calorimeter will greatly enhance the reconstruction efficiency as well.

The scatter plots in Fig. \ref{fig:omegaenergyscan} show the pseudorapidity distributions for the final-state photons in backward $\omega$ production at the three collision energies.  The top row shows the photons from the $\omega$ decay vs. photons from the $\pi^0$ decay, while the bottom row shows the two photons from the $\pi^0$. There is considerable correlation between the photon pseudorapidities.  

At 5$\times$41 GeV, the photon distribution is centered between the central and B0 detectors, with some acceptance in both.  As the energy rises, they shift to larger pseudorapidities.  At the highest energy, the ZDC becomes a reasonably efficient detector.  At all three energies, the detection efficiency rises significantly if a B0 calorimeter is present.

Table \ref{tab:geoeff} lists the geometric efficiencies for three proton beam energies, with and without a B0 calorimeter.  Because all three photons must be detected to reconstruct the $\omega$, the presence of the B0 calorimeter provides an order of magnitude improvement in the overall detection efficiency.  With it, backward $\omega$ production is observable at all three collision energies.  Without it, the efficiency is low except for 275 GeV protons.   From our simulations, we found that the electron beam energy does not have a significant effect on acceptance; the efficiency for 5 GeV electrons colliding with 100 GeV protons is very similar to that for 10 GeV electrons.

The actual efficiency is likely to be lower than the geometric efficiency due to numerous effects, such as particle interactions in the beampipe, photon merging in the calorimeter, finite resolution, and other phenomena.   However, the sum of these effects is expected to be relatively small, and should not change our overall conclusions.   

\subsection{$\rho^0\rightarrow\pi^+\pi^-$}
\label{subsection:rhoproduction}

The $\rho^0\rightarrow\pi^+\pi^-$ decay channel is an excellent option for measuring backward production, with a branching ratio of $\sim$100\% and only two charged pions in the final state.  

\begin{figure*}
  \begin{center}
    \includegraphics[width=1.0\textwidth]{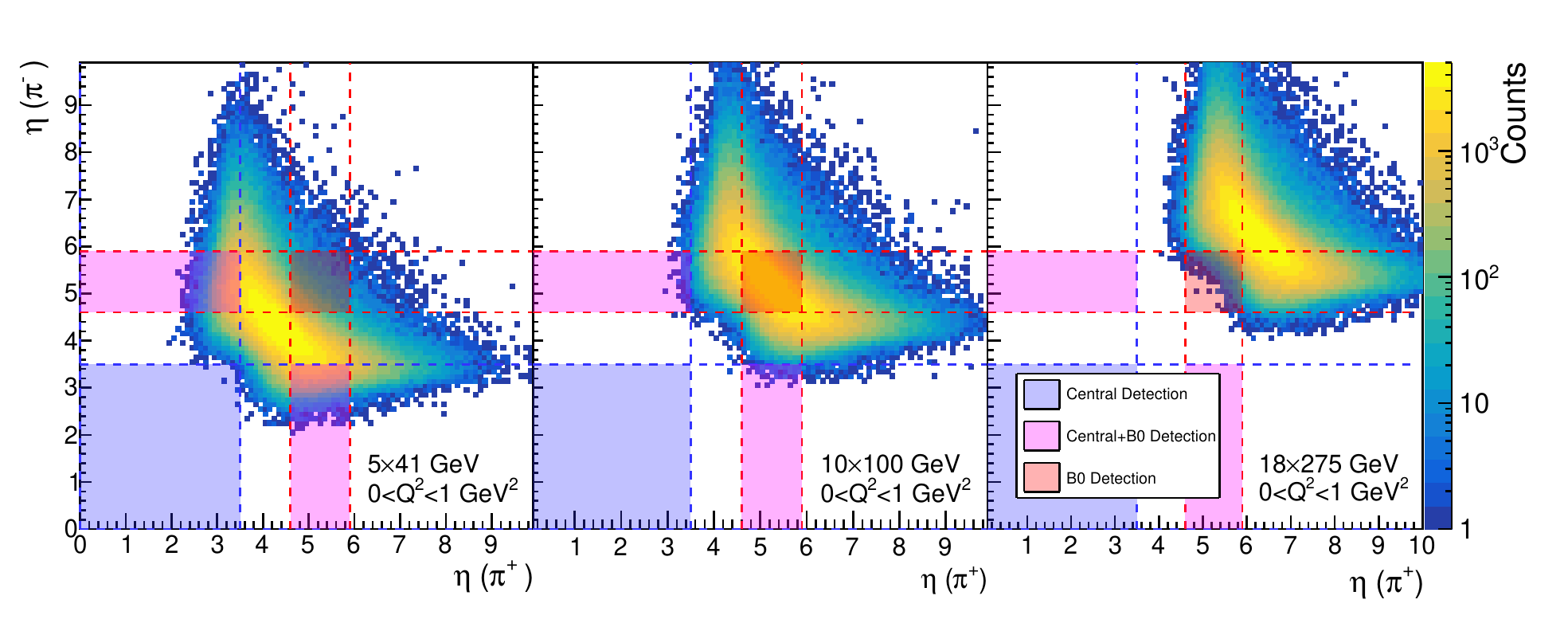}
\caption{Scatter plot of the pseudorapidity distribution of $\pi^{\pm}$ from $u$-channel $\rho^0$ production at three EIC energies.  The shaded colored regions show the acceptance in pseudorapidity of the central and B0 detectors.
    }
    \label{fig:rhoenergyscan}
  \end{center}
\end{figure*}

The pions can be detected only in the central and B0 detectors, so lower beam energies are likely to lead to higher acceptances.  Figure \ref{fig:rhoenergyscan} shows the joint pseudorapidity distributions of the two pions.  Even at the lowest energy of 5$\times$41 GeV, the vector-meson daughters are too far-forward for complete reconstruction within the central detectors. However, beam energies of 10$\times$100 GeV result in a two-pion distribution that is peaked within the geometric acceptance of the B0.

The geometric efficiencies are  given in Tab. \ref{tab:geoeff}.  The efficiency is highest for 10$\times$100 GeV collisions, and acceptable, even at the lowest beam energy.  For 18$\times$275 GeV collisions, the practicality of detection will likely depend on the details of the detector design.

\subsection{Prospects for Backward Production of Other Mesons, and with Heavier Ions}

To fully characterize backward production, it is important to probe as many different mesons as possible, which will differ in rates and kinematic characteristics.   As we saw, the backward production rate for the $\omega$ was about 30 million events per year.  Mesons with a production rate 100-1,000 times lower would still be visible if they could be detected with good efficiency.  The reconstruction efficiency will depend on the rapidity distribution of the produced meson. The heavier the meson, the higher the photon energy threshold, and the lower the rapidity of the $\gamma p$ system.  The lower the system's rapidity, the lower the rapidity of the vector meson.

The rapidity of vector mesons originating from $u$-channel production can be roughly estimated by making a few reasonable approximations. The first approximation is that the vector meson carries away all the energy of the initial-state proton ($E_p$). The second approximation is that the mass of the vector meson is much less than its energy. In these limits,

\begin{equation}
    y_{\text{V}} \approx 0.7 - \text{ln}\bigg(\frac{m_V}{E_p}\bigg).
\end{equation}

From this equation the location of the rapidity peak of other mesons can be estimated. For example, at 5$\times$41 GeV the $\rho^0$ will have a peak near $y\approx4.7$ which is born out by Fig. \ref{fig:rapiditycomparison}. The success of the B0 charged-particle tracker in detecting the $\rho^0$ demonstrates that other vector mesons with charged-particle decays may be appealing subjects of study. 

Backward production of heavier mesons, like the $\phi$ and potentially $J/\psi$, is of particular interest, since they share no valence quarks with the proton.  This is likely to reduce the ratio of the backward to forward cross sections, compared to the $\rho$ and $\omega$.   However, unless this additional suppression is very large, the $\phi$ is likely to be visible.  The $\phi$ will have typical average rapidities of 6.3, 5.3 and 4.4 for 275 GeV, 100 GeV and 41 GeV protons.  This is outside the range of the envisioned central trackers, but the 100 GeV proton data should have the $\phi$ well centered in the B0 detector.  

In addition to sharing no valence quarks with the incident baryon, the $J/\psi$ is composed of heavy charm quarks, so it is likely even further suppressed.  It will have typical average rapidities of 5.2, 4.2 and 3.3 at the three proton energies.  The 275 GeV proton data will produce $J/\psi$ that are visible in the B0 spectrometer, while backward $J/\psi$ from 41 GeV protons may be visible in the central detector, although with limited efficiency.   The $J/\psi \rightarrow e^+e^-$ channel will be measurable in both the charged-particle trackers and the electromagnetic calorimeters. 

The $\pi^0$ is also of interest at the EIC.   Even with 41 GeV protons, backward-produced $\pi^0$ will have an average rapidity near 6.3, and thus will only be visible in the ZDC. Detailed studies indicate that the reference ZDC will be able to reconstruct $\pi^0\rightarrow\gamma\gamma$ with good efficiency, at least for 40-60 GeV/c $\pi^0$ \cite{Li:2021aei}, thus allowing for the study of electroproduction as well as photoproduction.  

Backward production may not be limited to mesons.  It will be interesting to search for backward deeply-virtual Compton Scattering and Timelike Compton Scattering at the EIC \cite{Gayoso:2021rzj}.

So far, backward production has largely been studied with proton targets. However, heavier targets are also of interest.  It should be possible to use a deuterium beam, with the proton tagged in a forward spectrometer, to study backward production on neutron targets.  It would also allow for the study of production using heavy nuclei, to look for possible nuclear modifications to the cross section.   

\section{Backward Production at Other Colliders}

At the proposed Chinese EiCC, the 20 GeV proton beams should lead to backward vector-meson production that is shifted by about 0.7 units of rapidity compared to the 41 GeV proton rapidities listed above.  The detector design for the EiCC is at an early stage of development, but it is likely that most of these channels can be studied with a typical full-acceptance detector.  The energy range probed by the EiCC would complement the energies available at the EIC.

At the LHeC, with 7 TeV protons, the typical light-meson backward-production rapidities are around 10.0.  This is likely to be very difficult to study, beyond the capabilities of even fine-grained zero degree calorimeters. 

\section{Conclusions}

Baryon stopping can be profitably studied at the proposed U.S. electron-ion collider. These reactions lead to a final state that is comprised of a proton at mid-rapidity and a forward vector meson.  At the EIC, the $\omega$ rates range from 5 million to 100 million events/year, depending on beam energy and accelerator configuration - enough for detailed study.  The vector meson is produced at forward rapidity, and can, depending on its final state, be detected by the proposed EIC detectors. 

Backward production shares a deep theoretical connection with the baryon-junction model that may explain baryon stopping in heavy-ion collisions.  

We acknowledge useful conversations with Bill Li, Alex Jentsch, Prithwish Tribedy and Zhangbu Xu. Aaron Stanek participated in an early phase of this work. 
This work is supported in part by the U.S. Department of Energy, Office of Science, Office of Nuclear Physics, under contract numbers DE-AC02-05CH11231, by the US National Science Foundation under Grant
No. PHY-1812398, and by the University of California Office of the President Multicampus Research Programs under grant number M21PR3502.  

\bibliographystyle{apsrev4-1} 
\bibliography{main}
\end{document}